\documentclass[lettersize,journal]{IEEEtran}
\usepackage{amsmath,amsfonts}
\usepackage{amssymb}
\usepackage{algorithmic}
\usepackage{array}
\usepackage[caption=false,font=normalsize,labelfont=sf,textfont=sf]{subfig}
\usepackage{textcomp}
\usepackage{stfloats}
\usepackage{url}
\usepackage{verbatim}
\usepackage{graphicx}
\usepackage{booktabs}
\usepackage{makecell} 
\usepackage{tabularx} 
\usepackage{bm}
\usepackage{grffile}
\usepackage{color,xcolor}
\usepackage{caption}
\usepackage{epstopdf}
\def\BibTeX{{\rm B\kern-.05em{\sc i\kern-.025em b}\kern-.08em
    T\kern-.1667em\lower.7ex\hbox{E}\kern-.125emX}}
\usepackage{balance}
\begin{document}
\title{Resource Management for IRS-Assisted Full-Duplex
	Integrated Sensing, Communication and Computing Systems}
\author{Wanming Hao,~\IEEEmembership{Senior Member,~IEEE,} Xue Wu, Xingwang Li,~\IEEEmembership{Senior Member,~IEEE,} Gangcan Sun, Qingqing Wu ~\IEEEmembership{Senior Member,~IEEE}, Liang Yang ~\IEEEmembership{Senior Member,~IEEE} 
	\thanks{W. Hao, X. Wu, and G. Sun are with School of Electrical and Information Engineering, Zhengzhou University, Zhengzhou 450001, China. (e-mail: iewmhao@zzu.edu.cn, wuxue@gs.zzu.edu.cn, iegcsun@zzu.edu.cn)}
	\thanks{X. Li is with the School of Physics and Electronic Information Engineering, Henan Polytechnic University, Jiaozuo 454003, China. (email: lixingwang@hpu.edu.cn)}
	\thanks{Q. Wu is  with 
		the Department of Electronic Engineering, Shanghai Jiao Tong
		University, Shanghai 200240, China. (e-mail: qingqingwu@sjtu.edu.cn)} 
	\thanks{L. Yang is with College of Computer Science and Electronic Engineering, Hunan University, Changsha 410082, China. (e-mail: liangy@hnu.edu.cn)}
}
\maketitle

\begin{abstract}
In this paper, we investigate an intelligent reflecting surface (IRS) assisted full-duplex (FD) integrated sensing, communication and computing system. Specifically, an FD base station (BS) provides service for uplink and downlink transmission, and a local cache is connected to the BS through a backhaul link to store data. Meanwhile, active sensing elements are deployed on the IRS to receive target echo signals. On this basis, in order to evaluate the overall performance of the system under consideration, we propose a system utility maximization problem while ensuring the sensing quality, expressed as the difference between the sum of communication throughput, total computation bits (offloading bits and local computation bits) and the total backhaul cost for content delivery. This makes the problem difficult to solve due to the highly non-convex coupling of the optimization variables. To effectively solve this problem, we first design the most effective caching strategy. Then, we develop an algorithm based on weighted minimum mean square error, alternative direction method of multipliers, majorization-minimization framework, semi-definite relaxation techniques, and several complex transformations to jointly solve the optimization variables. Finally, simulation results are provided to verify the utility performance of the proposed algorithm and demonstrate the advantages of the proposed scheme compared with the baseline scheme.

\end{abstract}


\section{Introduction}
With the rapid development of future wireless systems, integrated sensing and communication (ISAC) technology has become a research hotspot in academia and industry \cite{ref1}-\cite{ref3}. By enabling wireless communication and radar sensing to use the same frequency, radio signals and infrastructure at ISAC system, greatly improving the spectral efficiency \cite{ref4}. At the same time, the application of massive terahertz communication technologies, multiple-input multiple-output (MIMO) and millimeter wave in radar sensing has further promoted the realization of ISAC technology, enabling ISAC to have higher resolution in many important application scenarios \cite{ref4-1}. Therefore, ISAC is considered to have great development prospects in future wireless networks. However, to cope with the growing computing demands of intelligent applications, more intelligent algorithms are proposed for data processing, which inevitably turns into a heavy computing load \cite{ref4-2}. To effectively solve these problems, a widely acclaimed solution is mobile edge computing and caching (MECC) \cite{ref5}, \cite{ref6}. By partially or completely offloading computation datas to the network edge and utilizing the local cache connected to the MEC server to prefetch user data, a higher network performance can be achieved as well as user service quality can be improved \cite{ref7}, \cite{ref8}.

Due to the complex wireless propagation environment, network performance will be significantly degraded when the direct link between the BS and user is blocked by obstacles such as buildings. In order to solve this issue, reconfigurable intelligent surface (RIS) or intelligent reflecting surface (IRS) has become another key technology in future wireless system \cite{ref9}, \cite{ref9-1}. IRS composed of a large number of low-cost passive components can provide virtual line-of-sight (LoS) links to bypass obstacles and expand wireless coverage, thereby improving system performance accordingly \cite{ref10}. In addition, IRS can reconfigure the wireless environment by adjusting its phase shift to mitigate interference, enhance the received signal strength \cite{ref11}, reduce transmission power consumption and hardware cost \cite{ref12}, and improve transmission reliability \cite{ref13}.

\subsection{Related Works}

In recent years, considering the advantages of combining ISAC and MEC technologies, researchers have shown increasing interest for integrated sensing, communication and computing (ISCC). Specifically, the authors of \cite{ref14} studied an integrated communication, radar sensing and edge computing network. By allocating spectrum and time resources, they proposed a standardized utility function maximization problem consisting of computing task delay and communication performance while ensuring sensing performance. The authors of \cite{ref15} combined the MEC paradigm with ISAC, formulated a joint device association and subchannel allocation problem, and developed an iterative matching algorithm by introducing the blocking pair form. The authors of \cite{ref16} proposed the use of MIMO array and dual-function radar communication technology in the user terminal, and proposed a multi-objective optimization (MOO) problem by jointly considering computation offloading and radar sensing. The work of \cite{ref17} adopted non-orthogonal multiple access (NOMA) technology and multi-layer computing structure, considers partial offloading and binary offloading at the same time, and jointly optimizes the transmit beamformer and the computing resources to maximize the computation offloading capability. The authors of \cite{ref18} studied an ISCC beyond fifth generation (B5G) cellular internet of thing (IoT) model and proposed two optimization problems: maximizing the weighted sum and minimizing the computation error by utilizing a new wireless computing technology. Different from traditional static sensors, the authors of \cite{ref19} used mobile crowd sensing (MCS) and proposed a joint sensing, communication, and computing (JSCC) framework for multi-dimensional resource constraints, which jointly optimized the sensing, computing, and transmission strategies to maximize the number of bits processed by the system.

Recently, due to the significant advantages of IRS, more studies have applied IRS to wireless communication sensing systems. Currently, IRS-enabled sensing generally adopts two methods: fully passive \cite{ref20}-\cite{ref22} and semi-passive \cite{ref23}-\cite{ref25}. For the semi-passive approach, active sensing elements are deployed on the IRS to receive and process the target echo signals. For the fully passive approach, there are no active sensing elements on the IRS, thus the echo signal is reflected back to the BS to perform target sensing. In \cite{ref20}, the authors considered the joint communication and sensing beamforming design and studied a RIS-assisted half-duplex (HD) ISAC system. By jointly optimizing user power, RIS phase shift, transmit and receive beamforming, they proposed a sum rate maximization problem. In \cite{ref21}, the authors estimated the angles and target response matrices of point targets and extended targets relative to the IRS based on the echo signal received by the BS, respectively, and then minimized the Cramer-Rao bound (CRB) by jointly optimizing the transmit and reflect beamforming. Since the introduced sensing function leads to the limitation of multi-user interference (MUI) on communication performance, the authors of \cite{ref22} minimized MUI under the CRB constraint of multi-target angle estimation by jointly optimizing the constant modulus waveform and discrete RIS phase shift. Different from the above-mentioned fully passive methods, in order to overcome the attenuation caused by IRS reflection, the authors of \cite{ref23} used an active IRS with an amplifier to estimate the angle of the point target based on the echo signal received on the IRS. In \cite{ref24}, the authors considered a multi-IRS-assisted ISAC system and proposed the problem of minimizing the CRB on all IRSs for point targets and extended targets by jointly optimizing the transmit and reflect beamforming. Then, the authors of \cite{ref25} compared the signal-to-noise ratio (SNR) between fully passive IRS and semi-passive IRS by jointly optimizing the transmit and reflected beamforming.

\subsection{Our Contributions}
As shown above, there are several papers that have conducted research on ISCC and semi-passive IRS-assisted ISAC systems. When studying the waveform design of ISCC or ISAC systems, most works considered half-duplex (HD) systems \cite{ref17}-\cite{ref19}, \cite{ref21}-\cite{ref22}. In the HD system, uplink reception and downlink transmission are operated separately in orthogonal frequency bands, which cannot fully utilize spectrum resources. In contrast, FD systems can achieve higher spectral efficiency \cite{ref20}, while how to solve the self-interference (SI) is difficult. More importantly, to the best of our knowledge, existing works have not considered deploying semi-passive IRS in FD ISCC. Based on the above discussion, we study an advanced semi-passive IRS-assisted FD ISCC system. The main contributions of this paper are elaborated as follows:
 
\begin{itemize}
	\item{In this paper, we integrate the semi-passive IRS into the FD ISCC system and study the corresponding resource allocation. In this system, the FD BS transmits the downlink synaesthesia signal while receiving the uplink offloading signal and the echo. At the same time, the FD BS owns a local cache for data storage.}
	\item{Our goal is to maximize the system utility, defined as the difference between the sum of communication throughput, total computation bits and the total backhaul cost for content delivery, by jointly optimizing the receive beamforming, transmit beamforming, IRS reflection coefficients, local computation resources, computation user power, and caching probability. Due to the existence of BS SI in FD systems and the impact of uplink offloading on co-channel interference (CCI) of downlink transmission, the optimization variables are highly non-convexly coupled, which cannot be directly addressed.}
	\item{To deal with the formulated problem, we first optimize the cache strategy and then reformulate the system utility maximization problem as a problem of maximizing the sum of communication throughput and total computation bits. In order to effectively solve the problem, we apply the weighted minimum mean square error (WMMSE) method to introduce auxiliary variables to reformulate the problem, then decompose it into several sub-problems, and transform each sub-problem into a convex problem through alternative direction method of multipliers (ADMM), majorization-minimization (MM), and semi-definite relaxation (SDR) techniques. Finally, the block coordinate ascent (BCA) method is used to alternately solve each optimization variable.}
	\item{Finally, the simulation results are provided to demonstrate the effectiveness of our proposed scheme. The results show that the proposed scheme owns higher system performance than other schemes.}
\end{itemize}

The rest of this paper is organized as follows: In section II, we propose a semi-passive IRS-assisted FD ISCC system and formulate a problem of maximizing the system utility function. Section III proposes a joint optimization framework. In Section IV, simulation results are presented to demonstrate the performance of the proposed scheme. Finally, the paper is concluded in Section V.

\textit{Notation:} In this paper, the matrices, vectors and scalars are denoted by the boldface uppercase boldface lowercase and lower-case letters, respectively; ${{\mathbf{A}}^{T}},{{\mathbf{A}}^{H}},\text{Tr}(\mathbf{A}),\text{Rank}(\mathbf{A})$ represent the transpose, conjugate transpose, trace and rank of matrix $\mathbf{A}$ respectively. $\mathbf{A }\text{ }\underline{\succ }\text{ }0 $ means that the matrix $\mathbf{A}$ is a semi-positive definite matrix. For a vector $\mathbf{a}$, ${{\left[ \mathbf{a} \right]}_{i}}$ means $i$-th elements, $\left\| \mathbf{a} \right\|$ represents its Euclidean norm, $\left| \mathbf{a} \right|$ represents its modulus; $\mathbb{E}\left[ x \right]$ represents the expectation of $x$. ${{\mathbf{I}}_{N}}$ represents an identity matrix of $N\times N$. $\mathcal{C}\mathcal{N}\left( \mu ,\sigma {}^{2} \right)$ represents a complex Gaussian random variable distribution with mean $\mu $ and variance ${{\sigma }^{2}}$. ${{\mathbb{C}}^{M\times N}}$ denotes a complex-valued matrix of size M $\times $ N.

\begin{figure}[!t]  
	\centering
	\includegraphics[width=3.5in]{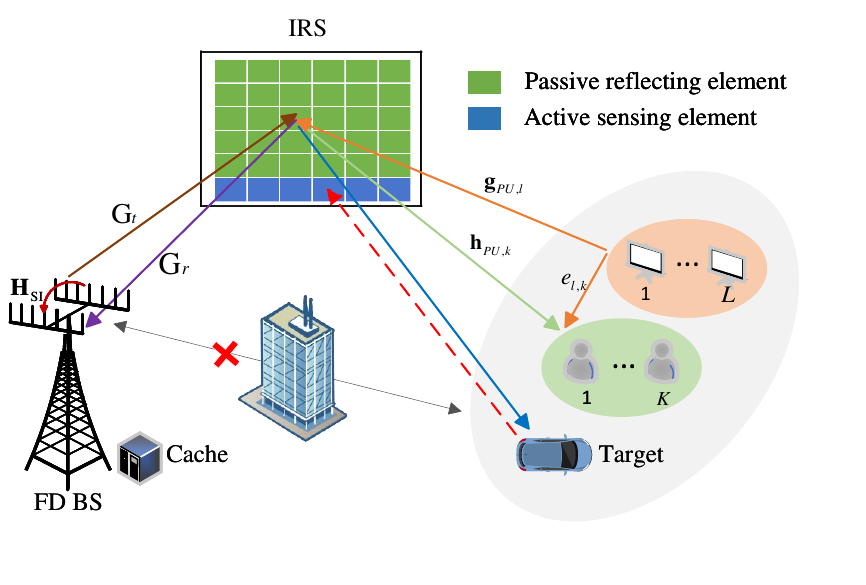}
	\captionsetup{justification=centering} 
	\caption{An IRS-assisted FD ISCC system.}
	\label{fig.1} 
\end{figure} 

\section{System Model}
As shown in Fig. 1, an IRS-assisted FD ISCC system is considered, where a FD BS is equipped with two uniform linear arrays (ULAs), a semi-passive IRS is equipped with ULAs consisting of $M$ passive reflection units, $K$ randomly distributed communication users (CM-UEs), $L$ randomly distributed computation users (CP-UEs), and a point sensing target. The FD BS is equipped with ${{N}_{t}}$ transmit antennas (TX) and ${{N}_{r}}$ receive antennas (RX), while each CM-UE and CP-UE is equipped with a single antenna. The FD BS simultaneously receives uplink offloading signals and transmits downlink ISAC signals for multi-user communication and target sensing with the help of IRS. In addition, the FD BS is connected to a local cache with limited storage capacity and connected to the MEC server through a high-capacity backhaul link, which is used for edge computing to perform computing tasks and store user data. For convenience, let $\mathcal{M} = \left\{ 1,\ldots ,M \right\}$, $\mathcal{K} = \left\{ 1,\ldots ,K \right\}$ and $\mathcal{L} = \left\{ 1,\ldots ,L \right\}$ denote the index of IRS elements, CM-UEs, CP-UEs. To address the severe path loss, we assume that in addition to $M$ passive reflecting elements (REs) in the semi-passive IRS, there are ${{M}_{a}}$ active sensing elements (SEs) for receiving the target echo signal. Furthermore, due to the unfavorable environmental propagation, we assume that there are obstacles between the BS and users as well as the target, and thus the direct links are blocked.

\subsection{Signal Model}

In this subsection, we first consider the downlink synaesthesia of the system, where a dual-function ISAC signals ${{\mathbf{x}}^{DL}}\in {{\mathbb{C}}^{{{N}_{t}}\times 1}}$ is used for simultaneous downlink multi-user communications and target sensing via multi-antenna beamforming. Let $s_{k}^{dl},k\in \left\{ 1,\cdots ,K \right\}$ denote the data symbol of the user $k$, and ${{\mathbf{w}}_{k}}\in {{\mathbb{C}}^{{{N}_{t}}\times 1}}$ denote the corresponding transmit beamforming vector. Similarly, ${{s}_{0}}$ represents the dedicated sensing signal and ${{\mathbf{w}}_{0}}\in {{\mathbb{C}}^{{{N}_{t}}\times 1}}$ represents the corresponding sensing wave beamforming. Assume that the transmitted information signals $s_{k}^{dl}$ and ${{s}_{0}}$ are both independent random variables with unit variance and mean 0, and that the downlink communication and target sensing signals are independent and uncorrelated, i.e., $\mathbb{E}[s_{k}^{dl}{{s}_{0}}]\text{ = }0,\forall k$. According to \cite{ref26}, the ISAC signal transmited by BS can be expressed as
\begin{equation}
{{\mathbf{x}}^{DL}}=\sum\limits_{k=1}^{K}{{{\mathbf{w}}_{k}}s_{k}^{dl}}+{{\mathbf{w}}_{0}}{{s}_{0}}.\tag{1} 
\end{equation}
In addition, we consider the BS total transmission power constraint as $\sum\nolimits_{k=0}^{K}{\left\| {{\mathbf{w}}_{k}} \right\|_{2}^{2}}\text{ }\le {{P}_{BS}}$, where ${{P}_{BS}}$ represents the maximum transmission power budget at the BS.

The FD BS simultaneously receives the target echo and the uplink offloading signal when it performs downlink transmission. The offloading signal transmitted by the $l$-th CP-UE can be written as
\begin{equation}
x_{l}^{UL}=\sqrt{{{p}_{l}}}s_{l}^{ul},\forall l\in \mathcal{L},\tag{2} 
\end{equation}
where $s_{l}^{ul}$ and $p_{l}$ represent the data symbol and transmit power of the $l$-th CP-UE, respectively. For simplicity, we assume that each $s_{l}^{ul}$ has unit variance and zero mean, and is uncorrelated with each other.

\subsection{Communication Model}

The received signal at CM-UE $k$ can be expressed as
\begin{align*}
{{y}_{com,k}} =&\underbrace{\mathbf{h}_{PU,k}^{H}\mathbf{\Phi }{{\mathbf{G}}_{t}}{{\mathbf{w}}_{k}}s_{k}^{dl}}_{\text{Desired  signal}}\tag{3}\\ 
&+\underbrace{\sum\limits_{k'\ne k}^{K}{\mathbf{h}_{PU,k}^{H}\mathbf{\Phi }{{\mathbf{G}}_{t}}{{\mathbf{w}}_{k'}}s_{k'}^{dl}}}_{\text{Multiuser interference}}+\underbrace{\mathbf{h}_{PU,k}^{H}\mathbf{\Phi }{{\mathbf{G}}_{t}}{{\mathbf{w}}_{0}}{{s}_{0}}}_{\text{Sensing signal}} \\ 
&+\underbrace{\sum\limits_{l=1}^{L}{\left( {{e}_{l,k}}+\mathbf{h}_{PU,k}^{H}\mathbf{\Phi }{{\mathbf{g}}_{PU,l}} \right)\sqrt{{{p}_{l}}}s_{l}^{ul}}}_{\text{UL interference and their reflections from IRS}}+{{n}_{k}}, 
\end{align*}
where ${{\mathbf{h}}_{PU,k}}\in {{\mathbb{C}}^{M\times 1}}$ and ${{\mathbf{G}}_{t}}\in {{\mathbb{C}}^{M\times {{N}_{t}}}}$ represent the channel vectors between the CM-UE $k$ and IRS, and between BS TX and IRS, respectively. ${{e}_{l,k}}\in \mathbb{C}$ and ${{\mathbf{g}}_{PU,l}}\in {{\mathbb{C}}^{M\times 1}}$ represent the channels between the CM-UE $k$ and the CP-UE $l$, and between the CP-UE $l$ and the IRS, respectively. 
${{n}_{k}}\sim\mathcal{C}\mathcal{N}\left( 0,\sigma _{k}^{2} \right)$ represents the additive white Gaussian noise (AWGN) at the user receiver. The diagonal matrix $\mathbf{\Phi }=\text{diag}\left( {{e}^{j{{\phi }_{1}}}},\cdots ,{{e}^{j{{\phi }_{M}}}} \right)$ represents the phase shift matrix of IRS, where ${{\phi }_{m}}\in [0,2\pi ),\forall m\in \mathcal{M}$ represents the phase shift of the $m$-th reflection element.

Therefore, the signal-to-interference plus noise-ratio (SINR) of the CM-UE $k$ can be written as
\begin{equation}
r_{k}^{com}=\frac{{{\left| {{\mathbf{h}}_{k}}{{\mathbf{w}}_{k}} \right|}^{2}}}{\sum\limits_{k'=0,k'\ne k}^{K}{{{\left| {{\mathbf{h}}_{k}}{{\mathbf{w}}_{k'}} \right|}^{2}}}+\sum\limits_{l=1}^{L}{{{p}_{l}}{{\left| {{{\bar{e}}}_{l,k}} \right|}^{2}}}+\sigma _{k}^{2}},\tag{4}
\end{equation}
where ${{\mathbf{h}}_{k}}\triangleq \mathbf{h}_{PU,k}^{H}\mathbf{\Phi }{{\mathbf{G}}_{t}}$ and ${{\bar{e}}_{l,k}}\triangleq {{e}_{l,k}}+\mathbf{h}_{PU,k}^{H}\mathbf{\Phi }{{\mathbf{g}}_{PU,l}}$.

The achievable communication rate from BS to CM-UE $k$ is given as
\begin{equation}
R_{k}^{com}=B\log \left( 1+r_{k}^{com} \right),\forall k\in \mathcal{K},\tag{5}
\end{equation}
where $B$ represents the transmission bandwidth.

\subsection{Sensing Model}
Next, we model the target echo signal, assuming that the radar signal composes of a non-line-of-sight path (NLoS), and the receive ULA at the IRS and the transmit ULA at the BS are both half-wavelength antenna spacing. The signal received on the IRS SE is expressed as
\begin{align*}
{{\mathbf{y}}_{IRS}}=&\underbrace{{{\eta }_{RT}}{{\mathbf{a}}_{a}}\left( \theta  \right)\mathbf{a}_{p}^{H}\left( \theta  \right)\mathbf{\Phi }{{\mathbf{G}}_{t}}{{\mathbf{x}}^{DL}}}_{\text{Desired echo signal}}\\
&+\underbrace{\sum\limits_{l=1}^{L}{\sqrt{{{p}_{l}}}{{\mathbf{g}}_{AU,l}}s_{l}^{ul}}}_{\text{UL interference signals}}+{{\mathbf{n}}_{s}},\tag{6}
\end{align*}
where ${{\eta }_{RT}}\in \mathbb{C}$ represents the reflection coefficient of the target, which is mainly determined by the radar cross section and path loss \cite{ref17}. $\theta$ represents the position of the sensing target with respect to the IRS. And ${{\mathbf{n}}_{s}}\sim\mathcal{C}\mathcal{N}\left( 0,\sigma _{s}^{2}{{\mathbf{I}}_{Ma}} \right)$ represents the AWGN at the IRS. ${{\mathbf{a}}_{a}}\left( \theta  \right)={{[1,{{e}^{-j2\pi \frac{d}{\lambda }\sin \theta }},\cdots ,{{e}^{-j2\pi \frac{d}{\lambda }\left( {{M}_{a}}-1 \right)\sin \theta }}]}^{T}}$ and ${{\mathbf{a}}_{p}}\left( \theta  \right)={{[1,{{e}^{-j2\pi \frac{d}{\lambda }\sin \theta }},\cdots ,{{e}^{-j2\pi \frac{d}{\lambda }\left( M-1 \right)\sin \theta }}]}^{T}}$ represent the receive steering vector of IRS SE and the transmit steering vector of IRS RE respectively. As described in \cite{ref27}, we assume that $\theta $ and ${{\eta }_{RT}}$ are known or previously estimated in order to design the most appropriate emission signal to detect this specific target.

The performance of radar is largely related to its corresponding SINR. Specifically, the detection probability of a point target usually increases with the increase of the output SINR in a MIMO radar system~\cite{ref28}. Therefore, we use radar SINR directly to characterize the sensing performance. According to (6), the radar SINR can be written as
\begin{equation}
{{r}^{tar}}=\frac{\left\| {{\eta }_{RT}}{{\mathbf{a}}_{a}}\left( \theta  \right)\mathbf{a}_{p}^{H}\left( \theta  \right)\mathbf{\Phi }{{\mathbf{G}}_{t}}{{\mathbf{x}}^{DL}} \right\|_{2}^{2}}{\sum\limits_{l=1}^{L}{{{p}_{l}}\left\| {{\mathbf{g}}_{AU,l}} \right\|_{2}^{2}}+\sigma _{s}^{2}}.\tag{7}
\end{equation}

\subsection{Computation Model}
\textit{1) Offloading computing:}

The offloading signal received by the FD BS can be expressed as
\begin{equation}
{{\mathbf{y}}_{BS}}=\underbrace{\sum\limits_{l=1}^{L}{\sqrt{{{p}_{l}}}\mathbf{G}_{r}^{H}\mathbf{\Phi }{{\mathbf{g}}_{PU,l}}s_{l}^{ul}}}_{\text{Offloading signal}}+\underbrace{{{\mathbf{H}}_{SI}}{{\mathbf{x}}^{DL}}}_{\text{SI}}+{{\mathbf{n}}_{b}},\tag{8}
\end{equation}
where ${{\mathbf{G}}_{r}}\in {{\mathbb{C}}^{M\times {{N}_{r}}}}$ represents the channel vector between BS RX and IRS. ${{\mathbf{n}}_{b}}\sim\mathcal{C}\mathcal{N}\left( 0,\sigma _{b}^{2}{{\mathbf{I}}_{{{N}_{r}}}} \right)$ represents the AWGN at the BS receiver. SI power can be greatly reduced by utilizing SI cancellation techniques. Without loss of generality, we represent the residual SI signal as ${{\mathbf{H}}_{SI}}{{\mathbf{x}}^{DL}}$ \cite{ref29}, where ${{\mathbf{H}}_{SI}}\in {{\mathbb{C}}^{{{N}_{r}}\times {{N}_{t}}}}$ represents the residual SI channel at the FD BS. Due to the severe attenuation of the channel between BS and IRS, we ignore the SI introduced by the IRS reflected synaesthesia signal and the reflected echo interference \cite{ref30}.

Technically, we apply a receive beamforming $\mathbf{u}_{l}^{H}\in {{\mathbb{C}}^{{{N}_{r}}\times 1}}$ at the BS to recover the offloading signal, and we obtain the corresponding receive SINR for the CP-UE $l$ as
\begin{equation}
r_{l}^{off}=\frac{{{p}_{l}}{{\left| \mathbf{u}_{l}^{H}{{\mathbf{g}}_{l}} \right|}^{2}}}{\sum\limits_{l'\ne l}^{L}{{{p}_{l'}}{{\left| \mathbf{u}_{l}^{H}{{\mathbf{g}}_{l'}} \right|}^{2}}+{{\left| \mathbf{u}_{l}^{H}{{\mathbf{H}}_{SI}}\mathbf{x} \right|}^{2}}+\left\| \mathbf{u}_{l}^{H} \right\|_{2}^{2}\sigma _{b}^{2}}},\tag{9}
\end{equation}
where ${{\mathbf{g}}_{l}}=\mathbf{G}_{r}^{H}\mathbf{\Phi }{{\mathbf{g}}_{PU,l}}$.

Therefore, the achievable offloading rate of the CP-UE $l$ at the BS is given as
\begin{equation}
R_{l}^{off}=B\log \left( 1+r_{l}^{off} \right),\forall l\in \mathcal{L}.\tag{10}
\end{equation}

\textit{2) Local computing:}

We assume that ${{\varepsilon }_{l}}$ represents the CPU cycle required for each CP-UE to process a bit of raw data, ${{f}_{l}}$ represents the local CPU computing frequency of the $m$-th CP-UE, and $\zeta$ represents the constant related to the hardware architecture. According to \cite{ref5}, \cite{ref30-1}, the computation rate and energy consumption at each CP-UE can be expressed as $R_{l}^{loc}={{f}_{l}}/{{\varepsilon }_{l}}$ and $E_{l}^{loc}=T\zeta f_{l}^{3}$, respectively.

\subsection{Cache Model}
Next, we model the caching strategy at the BS. The file collection $V$ is denoted by $\mathcal{V}=\{1,\cdots ,V\}$, and use vector $\mathbf{q}=[{{q}_{1}},\cdots {{q}_{V}}]$ to represent the length of these files. We assume that in the content database $V$, all files can be ranked according to their popularity. Furthermore, we utilize a probabilistic caching strategy to randomly cache content with deterministic probability. The caching strategy can be defined as $\mathbf{e}={{[{{e}_{1}},\cdots {{e}_{V}}]}^{T}}$, where ${{e}_{v}}\in [0,1],\forall v\in [1,V]$ represents the content placement indicator. In addition, since the BS cache is limited by the cache capacity $F$, there is a local cache constraint $\sum\limits_{v=1}^{V}{{{q}_{v}}{{e}_{v}}}\le F$. Furthermore, the probability of the $v$-th file requested by the user is denoted as ${{\tilde{c}}_{v}},\forall v\in [1,V]$. For simplicity, we assume that the request probability of the $v$-th file is consistent among different users without loss of generality, e.g., ${{\tilde{c}}_{v,1}}=,\cdots ,{{\tilde{c}}_{v,L}}\triangleq {{\tilde{c}}_{v}}$. It is assumed that users request content files based on their popularity, and the request probability can be modeled as a Zipf distribution, i.e., ${{\tilde{c}}_{v}}=\frac{{{v}^{-\epsilon }}}{\sum\nolimits_{i=1}^{V}{{{i}^{-\epsilon }}}}$, where $\epsilon $ is the skewness factor, which represents the deviation of the file popularity.

In content delivery, if the content $v$ is already cached in the local storage of the BS, users can directly access this content by requesting. Otherwise, we need to obtain the content from the BS via a backhaul link. We define the performance metric for acquiring content as the backhaul data rate, e.g., ${{R}_{0,l}}$. Therefore, within a channel coherence time $T$, the total backhaul cost can be modeled as ${{\mathcal{D}}_{total}}=T\sum\limits_{v=1}^{V}{{{\rho }_{v}}\sum\limits_{l=1}^{L}{\left( 1-{{e}_{v}} \right){{{\tilde{c}}}_{v}}{{R}_{0,l}}}}$, where ${{\rho }_{v}}$ represents the pricing of the backhaul cost for the $v$-th content delivery.

\begin{table}
	\begin{center}
		\begin{tabularx}{\linewidth}{l}
			\hline
			\toprule
			\textbf{Algorithm 1:} Proposed Algorithm for Solving ${{\mathcal{P}}^{0}}$\\
			\Xhline {1pt}	
			$\textbf{Input: }$${{\mathbf{G}}_{t}},{{\mathbf{G}}_{r}},{{\mathbf{G}}_{s}},{{\mathbf{H}}_{SI}},{{\mathbf{h}}_{PU,k}},{{\mathbf{g}}_{PU,l}},{{P}_{BS}},{{\Gamma }^{tar}},{{E}_{\max }}$ and $\left\{ {{R}_{0,l}} \right\}$.\\	
			$\textbf{Output: }$	$\mathbf{\hat{w}},\mathbf{\hat{u}},\boldsymbol{\hat{\phi }},\mathbf{\hat{f}},\mathbf{\hat{p}},\mathbf{\hat{e}}$.\\
			\text{ }1: Set maximum number of iterations: ${N}_{max}$ and $n$=0.\\
			\text{ }2: Find $\mathbf{e}$ by solving problem ${{\mathcal{P}}^{1}}$.\\
			\text{ }3: Initialize $\mathbf{w}$, $\boldsymbol{\phi }$, $\mathbf{f}$ and $\mathbf{p}$.\\
			\text{ }4: $\textbf{Repeat}$\\
			\text{ }5: \quad Upadate $\{{{\alpha }_{1,k}}\},\{{{\beta }_{1,k}}\}$ by (17) and (18).\\
			\text{ }6: \quad Upadate $\{{{\alpha }_{2,l}}\},\{{{\beta }_{2,l}}\}$ by (19) and (20).\\
			\text{ }7: \quad Upadate $\boldsymbol{\phi }$ by solving problem ${{\mathcal{P}}^{9}}$.\\
			\text{ }8: \quad Upadate $\boldsymbol{\psi }$ by (33).\\
			\text{ }9: \quad Upadate $\lambda$ by (34).\\
			10: \quad Upadate $\rho :=0.8\rho $.\\
			11: \quad Upadate $\mathbf{w}$ by solving problem ${{\mathcal{P}}^{13}}$.\\
			12: \quad Upadate $\mathbf{u}$ by (46).\\
			13: \quad Upadate $\mathbf{f}$ and $\mathbf{p}$ by solving problem ${{\mathcal{P}}^{18}}$.\\
			14: \quad Set $n$ = $n$ + 1.\\
			15: $\textbf{Until: }$$n$ = ${N}_{max}$ or the objective of problem  ${{\mathcal{P}}^{0}}$ converges.\\
			16: Return $\mathbf{\hat{w}}={{\mathbf{w}}^{(t+1)}},\mathbf{\hat{u}}={{\mathbf{u}}^{(t+1)}},\hat{\boldsymbol{\phi } }={{\boldsymbol{\phi } }^{(t+1)}},\mathbf{\hat{f}}={{\mathbf{f}}^{(t+1)}},$\\
			\quad\quad $\mathbf{\hat{p}}={{\mathbf{p}}^{(t+1)}}$ and $\mathbf{e}$.\\
			\Xhline {1pt}
		\end{tabularx}
	\end{center}
\end{table}

\subsection{Problem Formulation}
To evaluate the overall performance, we consider a system utility maximization problem, expressed as the gap between the sum of the communication throughput, the total computation bits (offloading bits and local computation bits) and the total backhaul cost for content delivery. Mathematically, the optimization problem can be expressed as follows
\begin{align*}
 {{\mathcal{P}}^{0}}:&\underset{\mathbf{w},\mathbf{u},\boldsymbol{\phi } ,\mathbf{f},\mathbf{p},\mathbf{e}}{\mathop{\max }}\,\text{    }\sum\limits_{k=1}^{K}{TR_{k}^{com}}+\text{ }\sum\limits_{l=1}^{L}{T\left( R_{l}^{off}+R_{l}^{loc} \right)}-{{\mathcal{D}}_{total}}\tag{11} \\ 
 \text{      }s.t.&\text{     }C1\text{ :  }\sum\limits_{k=0}^{K}{\left\| {{\mathbf{w}}_{k}} \right\|_{2}^{2}}\text{ }\le {{P}_{BS}},\tag{11a}  \\ 
& \text{     }C2\text{ :  }{{r}^{tar}}\ge {{\Gamma }^{tar}},\tag{11b}  \\ 
& \text{     }C3\text{ :  }\left| {{\phi }_{m}} \right|=1,\forall m\in \mathcal{M},\tag{11c}  \\ 
& \text{     }C4\text{ :  }T{{p}_{l}}+T\zeta f_{l}^{3}\le E_{l}^{\max },{{f}_{l}}\ge 0,{{p}_{l}}\ge 0,\tag{11d}  \\ 
& \text{     }C5\text{ :  }\sum\limits_{v=1}^{V}{{{q}_{v}}{{e}_{v}}}\le F,{{e}_{v}}\in [0,1],\tag{11e} 
\end{align*}
where ${{\Gamma }^{tar}}$ represents the target sensing SINR requirement, $E_{l}^{\max }$ represents the maximum energy budget of the $l$-th CP-UE. $\mathbf{w}=\{{{\mathbf{w}}_{k}}\}_{k=0}^{K}$ and $\mathbf{u}=\{{{\mathbf{u}}_{l}}\}_{l=1}^{L}$ represents the set of transmit beamforming and receive beamforming at the BS, respectively. $\mathbf{f}={{[{{f}_{1}},\cdots ,{{f}_{L}}]}^{T}}$ epresents the local computing resource vector. $\boldsymbol{\phi } ={{\left[ {{\phi }_{1}},\cdots ,{{\phi }_{M}} \right]}^{T}}$ epresents the phase shift vector of the reflection elements. $\mathbf{p}={{[{{p}_{1}},\cdots ,{{p}_{L}}]}^{T}}$ represents the CP-UEs power vector.

It is obvious that the formulated problem is a highly non-convex optimization one and is difficult to solve due to the logarithmic nature of the objective function, the coupling of the optimization variables, the radar SNR constraint $C$2, and the unit modulus constraint $C$3. To address these difficulties, we propose to transform the problem into several tractable sub-problems using BCA, WMMSE, MM, ADMM and SDP methods and solve them iteratively.

\section{Proposed Joint Optimization Framework}
In this section, we first optimize the cache strategy and then propose the WMMSE method to solve the problem of maximizing the sum of computation bits and communication throughput. Next, we apply the BCA method to divide the coupled variables into several blocks for alternate optimization. The details are as follows.

\subsection{Optimal Caching Strategy}
To optimize the design of the cache strategy, we rewrite the problem ${{\mathcal{P}}^{0}}$ as follows
\begin{align*}
 {{\mathcal{P}}^{1}}:&\underset{\mathbf{e}}{\mathop{\min }}\,\text{  }\sum\limits_{v=1}^{V}{\left( 1-{{e}_{v}} \right){{{\tilde{c}}}_{v}}}\text{  }\tag{12}\\ 
& \text{ }s.t.\text{     }C\text{5 :  }\sum\limits_{v=1}^{V}{{{q}_{v}}{{e}_{v}}}\le F,{{e}_{v}}\in \left[ 0,1 \right].\tag{12a} 
\end{align*}

Problem ${{\mathcal{P}}^{1}}$ is convex with respect to the optimization variables $\mathbf{e}={{[{{e}_{1}},\cdots {{e}_{V}}]}^{T}}$, and it can be solved directly using KKT optimality conditions or a convex optimization solver \cite{ref31}.

\subsection{Maximize the Sum of Total Computation Bits and Communication Throughput}
After optimizing the optimal caching strategy, problem ${{\mathcal{P}}^{0}}$ can be reformulated as maximizing the sum of communication throughput and total computation bits with given the content placement. Hence, we rewrite the problem as
\begin{align*}
& {{\mathcal{P}}^{2}}:\underset{\mathbf{w},\mathbf{u},\boldsymbol{\phi } ,\mathbf{f},\mathbf{p}}{\mathop{\max }}\,\text{    }\sum\limits_{k=1}^{K}{TR_{k}^{com}}+\text{ }\sum\limits_{l=1}^{L}{T\left( R_{l}^{off}+R_{l}^{loc} \right)}\tag{13} \\ 
& \text{ }s.t.\text{     }C1,\text{     }C2,\text{     }C3,\text{     }C4,\tag{13a} 
\end{align*}

In order to make it easier to solve problem ${{\mathcal{P}}^{2}}$, the objective function is first transformed via the WMMSE method \cite{ref32}. Specifically, by introducing auxiliary variables $\{{{\alpha }_{1,k}}\},\{{{\alpha }_{2,l}}\}$ and $\{{{\beta }_{1,k}}\},\{{{\beta }_{2,l}}\}$, we write the original objective function equivalently into two variant forms (14) and (15) as shown at the bottom of the next page \cite{ref32}.
\newcounter{One}
\setcounter{One}{\value{equation}}
\setcounter{equation}{13} 
\begin{figure*}[hb]
	\hrulefill
	\begin{flalign*} 
	& \log \left( 1+r_{k}^{com} \right)=\\
	&	\underset{{{\alpha }_{1,k}}\ge 0}{\mathop{\max }}\,\underbrace{\log \left( 1+{{\alpha }_{1,k}} \right)-{{\alpha }_{1,k}}+2\sqrt{1+{{\alpha }_{1,k}}}\Re \left\{ \beta _{1,k}^{*}{{\mathbf{h}}_{k}}{{\mathbf{w}}_{k}} \right\}-{{\left| {{\beta }_{1,k}} \right|}^{2}}\left( \sum\limits_{k'=0}^{K}{{{\left| {{\mathbf{h}}_{k}}{{\mathbf{w}}_{k'}} \right|}^{2}}}+\sum\limits_{l=1}^{L}{{{p}_{l}}{{\left| {{{\bar{e}}}_{l,k}} \right|}^{2}}}+\sigma _{k}^{2} \right)}_{\tilde{R}_{k}^{com}},\tag{14}\\
	& \log \left( 1+r_{l}^{off} \right)=\\
	& \underset{{{\alpha }_{2,l}}}{\mathop{\max }}\,\underbrace{\log \left( 1+{{\alpha }_{2,l}} \right)-{{\alpha }_{2,l}}+2\sqrt{1+{{\alpha }_{2,l}}}\Re \left\{ \beta _{2,l}^{*}\sqrt{{{p}_{l}}}\mathbf{u}_{l}^{H}{{\mathbf{g}}_{l}} \right\}-{{\left| {{\beta }_{2,l}} \right|}^{2}}\left( \sum\limits_{l'=1}^{L}{{{p}_{{{l}'}}}{{\left| \mathbf{u}_{l}^{H}{{\mathbf{g}}_{l'}} \right|}^{2}}+{{\left| \mathbf{u}_{l}^{H}{{\mathbf{H}}_{SI}}\mathbf{x} \right|}^{2}}+\left\| \mathbf{u}_{l}^{H} \right\|_{2}^{2}\sigma _{b}^{2}} \right)}_{\tilde{R}_{k}^{off}}.\tag{15}
	&
	\end{flalign*}		
\end{figure*}
\setcounter{equation}{\value{One}}

Therefore, original problem ${{\mathcal{P}}^{2}}$ can be equivalently expressed as
\begin{align*} 
 {{\mathcal{P}}^{3}}:&\underset{\mathbf{w},\mathbf{u},\boldsymbol{\phi } ,\mathbf{f},\mathbf{p}}{\mathop{\max }}\,\text{    }\sum\limits_{k=1}^{K}{\tilde{R}_{k}^{com}}+\text{ }\sum\limits_{l=1}^{L}{\left( \tilde{R}_{l}^{off}+\frac{R_{l}^{loc}}{B} \right)}\tag{16} \\ 
& \text{ }s.t.\text{     }C1,\text{     }C2,\text{     }C3,\text{     }C4,\tag{16a} 
\end{align*} 

Next, we apply the BCA method to solve problem ${{\mathcal{P}}^{3}}$.

\subsection{Optimizing Auxiliary Variables}
Based on the above WMMSE transformation, when other optimization variables are fixed, the optimization of auxiliary variables is an unconstrained convex problem, and the optimal solution of auxiliary variables can be obtained as follows
\begin{align*}
& {{{\hat{\alpha }}}_{1,k}}=\frac{{{\left| {{\mathbf{h}}_{k}}{{\mathbf{w}}_{k}} \right|}^{2}}}{\sum\limits_{k'=0,k'\ne k}^{K}{{{\left| {{\mathbf{h}}_{k}}{{\mathbf{w}}_{k'}} \right|}^{2}}}+\sum\limits_{l=1}^{L}{{{p}_{l}}{{\left| {{{\bar{e}}}_{l,k}} \right|}^{2}}}+\sigma _{k}^{2}}, \tag{17}\\ 
& {{{\hat{\beta }}}_{1,k}}=\frac{\sqrt{1+{{\alpha }_{1,k}}}{{\mathbf{h}}_{k}}{{\mathbf{w}}_{k}}}{\sum\limits_{k'=0}^{K}{{{\left| {{\mathbf{h}}_{k}}{{\mathbf{w}}_{k'}} \right|}^{2}}}+\sum\limits_{l=1}^{L}{{{p}_{l}}{{\left| {{{\bar{e}}}_{l,k}} \right|}^{2}}}+\sigma _{k}^{2}},\tag{18} \\ 
& {{{\hat{\alpha }}}_{2,l}}=\frac{{{p}_{l}}{{\left| \mathbf{u}_{l}^{H}{{\mathbf{g}}_{l}} \right|}^{2}}}{\sum\limits_{l'=1,l'\ne l}^{L}{{{p}_{{{l}'}}}{{\left| \mathbf{u}_{l}^{H}{{\mathbf{g}}_{l'}} \right|}^{2}}+{{\left| \mathbf{u}_{l}^{H}{{\mathbf{H}}_{SI}}\mathbf{x} \right|}^{2}}+\left\| \mathbf{u}_{l}^{H} \right\|_{2}^{2}\sigma _{b}^{2}}},\tag{19} \\ 
& {{{\hat{\beta }}}_{2,l}}=\frac{\sqrt{1+{{\alpha }_{1,l}}}\sqrt{{{p}_{l}}}\mathbf{u}_{l}^{H}{{\mathbf{g}}_{l}}}{\sum\limits_{l'=1}^{L}{{{p}_{{{l}'}}}{{\left| \mathbf{u}_{l}^{H}{{\mathbf{g}}_{l'}} \right|}^{2}}+{{\left| \mathbf{u}_{l}^{H}{{\mathbf{H}}_{SI}}\mathbf{x} \right|}^{2}}+\left\| \mathbf{u}_{l}^{H} \right\|_{2}^{2}\sigma _{b}^{2}}}.\tag{20}
\end{align*}

\subsection{Optimizing Phase Shift $\boldsymbol{\phi }$}
Based on other given variables, we study the optimization of the IRS phase shift $\boldsymbol{\phi }$. Problem ${{\mathcal{P}}^{3}}$ is reformulated as
\begin{align*}
{{\mathcal{P}}^{4}}:& \underset{\boldsymbol{\phi } }{\mathop{\max }}\,\text{   }\sum\limits_{k=1}^{K}{\tilde{R}_{k}^{com}}+\text{ }\sum\limits_{l=1}^{L}{\tilde{R}_{l}^{off}} \tag{21}\\ 
 \text{ }s.t.&\text{     }C2\text{ :  }{{r}^{tar}}\ge {{\Gamma }^{tar}},\tag{21a} \\ 
&\text{     }C3\text{ :  }\left| {{\phi }_{m}} \right|=1,\forall m\in \mathcal{M}.\tag{21b} 
\end{align*}

By introducing some new coefficients, the objective function and constraint $C$2 are rewritten as
\begin{flalign*}
&\sum\limits_{k=1}^{K}{\tilde{R}_{k}^{com}}+\text{ }\sum\limits_{l=1}^{L}{\tilde{R}_{l}^{off}}=-{{\boldsymbol{\phi } }^{H}}{{\mathbf{T}}_{12}}\boldsymbol{\phi }+2\Re \{\mathbf{t}_{12}^{H}\boldsymbol{\phi } \}+{{b}_{12}},\tag{22}\\
&\overset{\scriptscriptstyle\frown}{C}2\text{ : }{{b}_{0}}-{{\boldsymbol{\phi } }^{H}}{{\mathbf{T}}_{0}}\boldsymbol{\phi } \le 0,\tag{23}
&
\end{flalign*}
where some of these coefficients are defined in (24) as shown at the top of the next page.
\newcounter{Two}
\setcounter{Two}{\value{equation}}
\setcounter{equation}{23} 
\begin{figure*}[ht]
	\begin{flalign*} 
	& {{b}_{1,k}}=\log \left( 1+{{\alpha }_{1,k}} \right)-{{\alpha }_{1,k}}-{{\left| {{\beta }_{1,k}} \right|}^{2}}\sigma _{k}^{2}-{{\left| {{\beta }_{1,k}} \right|}^{2}}\sum\limits_{l=1}^{L}{{{p}_{l}}{{\left| {{e}_{l,k}} \right|}^{2}}},{{\mathbf{E}}_{1,k}}=\text{diag}\left( {{\mathbf{G}}_{t}}{{\mathbf{w}}_{k}} \right),{{\mathbf{a}}_{1,k,l}}=\mathbf{h}_{PU,k}^{H}\text{diag}\left( {{\mathbf{g}}_{PU,l}} \right),\tag{24} \\ 
	& {{\mathbf{t}}_{1,1,k}}=2\sqrt{1+{{\alpha }_{1,k}}}{{\beta }_{1,k}}\mathbf{E}_{1,k}^{H}{{\mathbf{h}}_{PU,k}},{{\mathbf{t}}_{1,2,k}}=2{{\left| {{\beta }_{1,k}} \right|}^{2}}\sum\limits_{l=1}^{L}{{{p}_{l}}{{e}_{l,k}}\mathbf{a}_{1,k,l}^{H}},{{\mathbf{t}}_{1,k}}={{\mathbf{t}}_{1,1,k}}-{{\mathbf{t}}_{1,2,k}},{{\mathbf{T}}_{1,1,k}}={{\left| {{\beta }_{1,k}} \right|}^{2}}\sum\limits_{l=1}^{L}{{{p}_{l}}\mathbf{a}_{1,k,l}^{H}{{\mathbf{a}}_{1,k,l}}}, \\ 
	& {{\mathbf{T}}_{1,2,k}}={{\left| {{\beta }_{1,k}} \right|}^{2}}\left( \sum\limits_{k'=0}^{K}{\mathbf{E}_{1,k'}^{H}{{\mathbf{h}}_{PU,k}}\mathbf{h}_{PU,k}^{H}{{\mathbf{E}}_{1,k'}}} \right),{{\mathbf{T}}_{1,k}}={{\mathbf{T}}_{1,1,k}}+{{\mathbf{T}}_{1,2,k}},{{b}_{1}}=\sum\limits_{k=1}^{K}{{{b}_{1,k}}},{{\mathbf{t}}_{1}}=\sum\limits_{k=1}^{K}{{{\mathbf{t}}_{1,k}}},{{\mathbf{T}}_{1}}=\sum\limits_{k=1}^{K}{{{\mathbf{T}}_{1,k}}}, \\
	& {{b}_{2,l}}=\log \left( 1+{{\alpha }_{2,l}} \right)-{{\alpha }_{2,l}}-{{\left| {{\beta }_{2,l}} \right|}^{2}}\left( {{\left| \mathbf{u}_{l}^{H}{{\mathbf{H}}_{SI}}\mathbf{x} \right|}^{2}}+\left\| \mathbf{u}_{l}^{H} \right\|_{2}^{2}\sigma _{b}^{2} \right),{{\mathbf{a}}_{2,l}}=\mathbf{G}_{r}^{H}\text{diag}\left( {{\mathbf{g}}_{PU,l}} \right),{{\mathbf{t}}_{2,l}}=2\sqrt{1+{{\alpha }_{2,l}}}{{\beta }_{2,l}}\sqrt{{{p}_{l}}}\mathbf{a}_{2,l}^{H}{{\mathbf{u}}_{l}}, \\ 
	& {{\mathbf{T}}_{2,l}}={{c}_{2,l}}\sum\limits_{l'=1}^{L}{{{p}_{l'}}\mathbf{a}_{2,l'}^{H}{{\mathbf{u}}_{l}}\mathbf{u}_{l}^{H}{{\mathbf{a}}_{2,l'}}},{{b}_{2}}=\sum\limits_{l=1}^{L}{{{b}_{2,l}}},{{\mathbf{t}}_{2}}=\sum\limits_{l=1}^{L}{{{\mathbf{t}}_{2,l}}},{{\mathbf{T}}_{2}}=\sum\limits_{l=1}^{L}{{{\mathbf{T}}_{2,l}}},{{\mathbf{T}}_{12}}={{\mathbf{T}}_{1}}+{{\mathbf{T}}_{2}},{{\mathbf{t}}_{12}}={{\mathbf{t}}_{1}}+{{\mathbf{t}}_{2}},{{b}_{12}}={{b}_{1}}+{{b}_{2}},\\
	&{{b}_{0}}={{\Gamma }^{tar}}\left( \sum\limits_{l=1}^{L}{{{p}_{l}}\left\| {{\mathbf{g}}_{AU,l}} \right\|_{2}^{2}}+\sigma _{s}^{2} \right),{{\mathbf{G}}_{s}}={{\eta }_{RT}}{{\mathbf{a}}_{a}}\left( \theta  \right)\mathbf{a}_{p}^{H}\left( \theta  \right),{{\mathbf{t}}_{0}}={{\mathbf{G}}_{t}}\mathbf{x},{{\mathbf{T}}_{0,1}}={{\mathbf{G}}_{s}}\text{diag}\left( {{\mathbf{t}}_{0}} \right),{{\mathbf{T}}_{0}}=\mathbf{T}_{0,1}^{H}{{\mathbf{T}}_{0,1}}.
	&
	\end{flalign*}	
	\hrulefill	
\end{figure*}
\setcounter{equation}{\value{Two}}

According to (22) and (23), we rewrite the reflection phase shift optimization problem as
\begin{align*}
{{\mathcal{P}}^{5}}:& \underset{\boldsymbol{\phi } }{\mathop{\min }}\,\text{   }{{\boldsymbol{\phi } }^{H}}{{\mathbf{T}}_{12}}\boldsymbol{\phi } -2\Re \{\mathbf{t}_{12}^{H}\boldsymbol{\phi } \}-{{b}_{12}}\tag{25}  \\ 
& \text{ }s.t.\text{     }\overset{\scriptscriptstyle\frown}{C}2\text{ :  }{{b}_{0}}-{{\boldsymbol{\phi } }^{H}}{{\mathbf{T}}_{0}}\boldsymbol{\phi } \le 0,\tag{25a}  \\ 
& \text{           }C3\text{ :  }\left| {{\phi }_{m}} \right|=1,\forall m\in \mathcal{M}.\tag{25b}
\end{align*}

We apply ADMM to solve problem ${{\mathcal{P}}^{5}}$. Since $C$3 is a non-convex constant amplitude constraint, a copy $\Psi$ of $\Phi$ is introduced to align and decouple. Problem ${{\mathcal{P}}^{5}}$ can be equivalently expressed as
\begin{align*}
 {{\mathcal{P}}^{6}}:&\underset{\boldsymbol{\phi } ,\boldsymbol{\psi } }{\mathop{\min }}\,\text{   }{{\boldsymbol{\phi } }^{H}}{{\mathbf{T}}_{12}}\boldsymbol{\phi } -2\Re \{\mathbf{t}_{12}^{H}\boldsymbol{\phi } \}-{{b}_{12}}\tag{26} \\ 
 \text{ }s.t.&\text{     }\overset{\scriptscriptstyle\frown}{C}2\text{ :  }{{b}_{0}}-{{\boldsymbol{\phi } }^{H}}{{\mathbf{T}}_{0}}\boldsymbol{\phi } \le 0 ,\tag{26a}\\ 
& \text{           }C6\text{ :  }\boldsymbol{\phi } =\boldsymbol{\psi } ,\tag{26b} \\ 
& \text{           }C7\text{ :  }\left| {{\psi }_{m}} \right|=1,\forall m\in \mathcal{M}.\tag{26c}
\end{align*}

In order to solve problem ${{\mathcal{P}}^{6}}$, based on ADMM \cite{ref33}, we optimize its augmented Lagrangian (AL) problem and rewrite the problem as
\begin{align*}
{{\mathcal{P}}^{7}}:& \underset{\boldsymbol{\phi } ,\boldsymbol{\psi } ,\mathbf{\lambda} }{\mathop{\min }}\,\text{   }{{\phi }^{H}}{{\mathbf{T}}_{12}}\phi -2\Re \{\mathbf{t}_{12}^{H}\boldsymbol{\phi } \}-{{b}_{12}}+\frac{1}{2\rho }\left\| \boldsymbol{\phi } -\boldsymbol{\psi } +\rho \mathbf{\lambda } \right\|_{2}^{2} \tag{27} \\ 
 \text{ }s.t.&\text{     }\overset{\scriptscriptstyle\frown}{C}2\text{ :  }{{b}_{0}}-{{\boldsymbol{\phi } }^{H}}{{\mathbf{T}}_{0}}\boldsymbol{\phi } \le 0,\tag{27a}  \\ 
& \text{           }C7\text{ :  }\left| {{\psi }_{m}} \right|=1,\forall m\in \mathcal{M} ,\tag{27b} 
\end{align*}
where $\lambda$ represents the dual variable and $\rho >0$ represents the penalty coefficient. We solve problem ${{\mathcal{P}}^{7}}$ by updating each variable alternately.

\textit{1) Update $\boldsymbol{\phi }$}

With given $\boldsymbol{\psi }$ and $\lambda$, the problem can be simplified as
\begin{align*}
 {{\mathcal{P}}^{8}}:&\underset{\boldsymbol{\phi } }{\mathop{\min }}\,\text{   }{{\boldsymbol{\phi } }^{H}}{{\mathbf{T}}_{12}}\boldsymbol{\phi } -2\Re \{\mathbf{t}_{12}^{H}\boldsymbol{\phi } \}-{{b}_{12}}+\frac{1}{2\rho }\left\| \boldsymbol{\phi } -\boldsymbol{\psi } +\rho \mathbf{\lambda } \right\|_{2}^{2} \tag{28}\\ 
\text{ }s.t.& \text{     }\overset{\scriptscriptstyle\frown}{C}2\text{ :  }{{b}_{0}}-{{\boldsymbol{\phi } }^{H}}{{\mathbf{T}}_{0}}\boldsymbol{\phi } \le 0.\tag{28a}
\end{align*}
Due to the constraints $\overset{\scriptscriptstyle\frown}{C}2$, problem ${{\mathcal{P}}^{8}}$ is non-convex. We adopt the MM framework \cite{ref34} to convexify $\overset{\scriptscriptstyle\frown}{C}2$ by linearizing the convex term at point ${{\boldsymbol{\phi } }_{0}}$, which is given as
\begin{equation}
{{\boldsymbol{\phi } }^{H}}{{\mathbf{T}}_{0}}\boldsymbol{\phi } \ge 2\Re \left\{ {{\boldsymbol{\phi } }_{0}}^{H}{{\mathbf{T}}_{0}}\left( \boldsymbol{\phi } -{{\boldsymbol{\phi } }_{0}} \right) \right\}+{{\boldsymbol{\phi } }_{0}}^{H}{{\mathbf{T}}_{0}}{{\boldsymbol{\phi } }_{0}}.\tag{29}
\end{equation}

Next, we replace ${{\boldsymbol{\phi } }^{H}}{{\mathbf{T}}_{0}}\boldsymbol{\phi }$ in constraint $\overset{\scriptscriptstyle\frown}{C}2$ and problem ${{\mathcal{P}}^{8}}$ can be rewritten as
\begin{align*}
 &{{\mathcal{P}}^{9}}:\underset{\boldsymbol{\phi } }{\mathop{\min }}\,\text{   }{{\boldsymbol{\phi } }^{H}}{{\mathbf{T}}_{12}}\boldsymbol{\phi } -2\Re \{\mathbf{t}_{12}^{H}\boldsymbol{\phi } \}-{{b}_{12}}+\frac{1}{2\rho }\left\| \boldsymbol{\phi } -\boldsymbol{\psi } +\rho \mathbf{\lambda } \right\|_{2}^{2} \tag{30}\\ 
 &\text{ }s.t.\text{     }\overset{\scriptscriptstyle\smile}{C}2\text{ :  }-2\Re \left\{ {{\boldsymbol{\phi } }_{0}}^{H}{{\mathbf{T}}_{0}}\boldsymbol{\phi }  \right\}+{{\left( {{\boldsymbol{\phi } }_{0}}^{H}{{\mathbf{T}}_{0}}{{\boldsymbol{\phi } }_{0}} \right)}^{*}}+{{b}_{0}}\le 0.\tag{30a}
\end{align*}
Problem ${{\mathcal{P}}^{9}}$ is a typical second-order cone program (SOCP) and can be solved directly by existing convex optimization solvers, such as CVX \cite{ref31}.

\textit{2) Update $\boldsymbol{\psi }$}

With fixed $\boldsymbol{\phi }$ and $\lambda$, the problem can be simplified as
\begin{align*}
 {{\mathcal{P}}^{10}}:&\underset{\boldsymbol{\psi } }{\mathop{\min }}\,\text{   }\frac{1}{2\rho }\left\| \boldsymbol{\phi } -\boldsymbol{\psi } +\rho \mathbf{\lambda } \right\|_{2}^{2} \tag{31}\\ 
 \text{ }s.t.&\text{     }C7\text{ :  }\left| {{\psi }_{m}} \right|=1,\forall m\in \mathcal{M}.\tag{31a} 
\end{align*}

Since $\boldsymbol{\psi }$ owns a unit modulus term, the quadratic term in the objective function in terms of $\boldsymbol{\phi }$ is a constant, i.e., $\left\| \boldsymbol{\psi }  \right\|_{2}^{2}/(2\rho )=M/(2\rho )$. Therefore problem ${{\mathcal{P}}^{10}}$ can be simplified to
\begin{equation}
{{\mathcal{P}}^{11}}:\underset{\left| \psi  \right|={{1}_{M}}}{\mathop{\max }}\,\text{   }\Re \left\{ {{\left( \boldsymbol{\phi } +\rho \lambda  \right)}^{H}}\boldsymbol{\psi }  \right\}\tag{32}
\end{equation}

It can be seen that when all elements of $\boldsymbol{\psi }$ are aligned with the elements of the $\left( {{\rho }^{-1}}\boldsymbol{\phi } +\lambda  \right)$, problem ${{\mathcal{P}}^{11}}$ can be maximized. Therefore its optimal solution is expressed as:
\begin{equation}
\hat{\boldsymbol{\psi } }={{e}^{j\cdot \angle \left( \boldsymbol{\phi } +\rho \lambda  \right)}}.\tag{33}
\end{equation}

\textit{3) Update $\lambda$}

After obtaining $\boldsymbol{\phi }$ and $\boldsymbol{\psi }$, the dual variable $\lambda$ is updated via gradient ascent, as follows
\begin{equation}
 \lambda :=\lambda +{{\rho }^{-1}}\left( \boldsymbol{\phi } -\boldsymbol{\psi }  \right).\tag{34}
\end{equation}

\subsection{Optimizing BS beamforming ${{\mathbf{w}}_{k}}$}
Next, we study the optimization of the BS beamforming ${{\mathbf{w}}_{k}}$ when other variables are fixed. We reformulate the problem as
\begin{align*}
 {{\mathcal{P}}^{12}}:&\underset{\mathbf{w}}{\mathop{\max }}\,\text{   }\sum\limits_{k=1}^{K}{\tilde{R}_{k}^{com}}+\text{ }\sum\limits_{l=1}^{L}{\tilde{R}_{l}^{off}} \tag{35}\\ 
 \text{ }s.t.&\text{     }C\text{1 : }\sum\limits_{k=0}^{K}{\left\| {{\mathbf{w}}_{k}} \right\|_{2}^{2}}\text{ }\le {{P}_{BS}},\tag{35a} \\ 
& \text{          }C2\text{ :  }{{r}^{tar}}\ge {{\Gamma }^{tar}}.\tag{35b}
\end{align*}

The objective function and constraint $C$2 are rewritten by introducing several new coefficients as
\begin{align*}
&\tilde{R}_{k}^{com}={{b}_{3,k}}+\sqrt{1+{{\alpha }_{1,k}}}\text{Tr}\left( {{\Omega }_{k}}{{{\mathbf{\tilde{W}}}}_{k}} \right)\\
 &-{{\left| {{\beta }_{1,k}} \right|}^{2}}\sum\limits_{k'=0}^{K}{\text{Tr}\left( \mathbf{h}_{k}^{H}{{\mathbf{h}}_{k}}{{\mathbf{W}}_{k'}} \right)},\tag{36} \\ 
&\tilde{R}_{l}^{off}={{b}_{4,l}}-{{\left| {{\beta }_{2,l}} \right|}^{2}}\text{Tr}\left( {{\mathbf{Z}}_{l}}\sum\limits_{k=0}^{K}{{{\mathbf{W}}_{k}}} \right),\tag{37}\\
&\bar{C}2\text{ : Tr}\left( {{\Omega }_{0}}\sum\limits_{k=0}^{K}{{{\mathbf{W}}_{k}}} \right)\ge {{b}_{0}},\tag{38}
\end{align*}
where some of these coefficients are defined as follows
\begin{flalign*}
&{{b}_{3,k}}=\log \left( 1+{{\alpha }_{1,k}} \right)-{{\alpha }_{1,k}}+1 \tag{39}\\
&-{{\left| {{\beta }_{1,k}} \right|}^{2}}\left( \sum\limits_{l=1}^{L}{{{p}_{l}}{{\left| {{e}_{l,k}}+\mathbf{h}_{PU,k}^{H}\mathbf{\Phi }{{\mathbf{g}}_{PU,l}} \right|}^{2}}}+\sigma _{k}^{2} \right),\\
& {{b}_{4,l}}=\log \left( 1+{{\alpha }_{2,l}} \right)-{{\alpha }_{2,l}}\\
&+2\sqrt{1+{{a}_{2,l}}}\Re \left\{ \beta _{2,l}^{*}\sqrt{{{p}_{l}}}\mathbf{u}_{l}^{H}\mathbf{G}_{r}^{H}\mathbf{\Phi }{{\mathbf{g}}_{PU,l}} \right\} \\ 
& \text{         }-{{\left| {{\beta }_{1,l}} \right|}^{2}}\left( \sum\limits_{l'=1}^{L}{{{p}_{l}}{{\left| \mathbf{u}_{l}^{H}\mathbf{G}_{r}^{H}\mathbf{\Phi }{{\mathbf{g}}_{PU,l'}} \right|}^{2}}}+\left\| \mathbf{u}_{l}^{H} \right\|_{2}^{2}\sigma _{b}^{2} \right), \\ 
&{{\mathbf{W}}_{k}}={{\mathbf{w}}_{k}}\mathbf{w}_{k}^{H},{{\mathbf{\tilde{w}}}_{k}}=\left[ \begin{matrix}
{{\mathbf{w}}_{k}}  \\
1  \\
\end{matrix} \right],{{\mathbf{\tilde{W}}}_{k}}={{\mathbf{\tilde{w}}}_{k}}\mathbf{\tilde{w}}_{k}^{H},
&
\end{flalign*}
\begin{flalign*}
&{{\Omega }_{k}}=\left[ \begin{matrix}
0 & {{\left( \beta _{1,k}^{*}{{\mathbf{h}}_{k}} \right)}^{H}}  \\
\beta _{1,k}^{*}{{\mathbf{h}}_{k}} & 0  \\
\end{matrix} \right],{{\Omega }_{0}}={{\left( {{\mathbf{G}}_{s}}\mathbf{\Phi }{{\mathbf{G}}_{t}} \right)}^{H}}{{\mathbf{G}}_{s}}\mathbf{\Phi }{{\mathbf{G}}_{t}},\\
&{{\mathbf{H}}_{k}}=\mathbf{h}_{k}^{H}{{\mathbf{h}}_{k}},{{\mathbf{Z}}_{l}}=\mathbf{H}_{SI}^{H}{{\mathbf{u}}_{l}}\mathbf{u}_{l}^{H}{{\mathbf{H}}_{SI}}.
&
\end{flalign*}

According to the above transformation, problem ${{\mathcal{P}}^{12}}$ can be reformulated as:
\begin{align*}
 {{\mathcal{P}}^{13}}:&\underset{{{{\mathbf{\tilde{W}}}}_{k}}}{\mathop{\max }}\,\text{   }\sum\limits_{k=1}^{K}{\left( \sqrt{1+{{\alpha }_{1,k}}}\text{Tr}\left( {{\Omega }_{k}}{{{\mathbf{\tilde{W}}}}_{k}} \right)\right.} \\
& {\left.-{{\left| {{\beta }_{1,k}} \right|}^{2}}\sum\limits_{k'=0}^{K}{\text{Tr}\left( {{\mathbf{H}}_{k}}{{\mathbf{W}}_{k'}} \right)} \right)} \\ 
&-\sum\limits_{l=1}^{L}{\left( {{\left| {{\beta }_{1,l}} \right|}^{2}}\text{Tr}\left( {{\mathbf{Z}}_{L}}\sum\limits_{k=0}^{K}{{{\mathbf{W}}_{k}}} \right) \right)}\tag{40} \\ 
\text{ }s.t.& \text{     }C\text{1 : Tr}\left( \sum\limits_{k=0}^{K}{{{\mathbf{W}}_{k}}} \right)\text{ }\le {{P}_{BS}},\tag{40a}  \\ 
& \text{          }\bar{C}2\text{ :  Tr}\left( {{\Omega }_{0}}\sum\limits_{k=0}^{K}{{{\mathbf{W}}_{k}}} \right)\ge {{b}_{0}},\tag{40b} \\ 
& \text{          }C8\text{ :  }{{\left[ {{{\mathbf{\tilde{W}}}}_{k}} \right]}_{{{N}_{t}}+1,{{N}_{t}}+1}}=1,{{{\mathbf{\tilde{W}}}}_{k}}\underset{\scriptscriptstyle-}{\succ }0,k\in \left\{ 0,\mathcal{K} \right\},\tag{40c}\\ 
& \text{          }C9\text{ :  rank}\left( {{{\mathbf{\tilde{W}}}}_{k}} \right)=1.\tag{40d}
\end{align*}

Due to the rank-one constraint of C9, ${{\mathcal{P}}^{13}}$ is still non-convex, so we choose to use SDP technology to obtain the optimal solution of the problem by relaxing constraint C9. Finally, we directly solve it through a convex optimization solver and use the Gaussian randomization method to recover the corresponding rank-one solution \cite{ref34-1}.

\subsection{Optimizing Offloading Signal Receive Beamforming ${{\mathbf{u}}_{l}}$}
Next, we study the optimization of the offloading signal receive beamforming ${{\mathbf{u}}_{l}}$ when other variables are fixed. We reformulate the problem as
\begin{equation}
{{\mathcal{P}}^{14}}:\underset{\mathbf{u}}{\mathop{\max }}\,\text{   }\sum\limits_{l=1}^{L}{\tilde{R}_{l}^{off}}\tag{41}
\end{equation}

We rewrite the objective function as:
\begin{equation}
\tilde{R}_{l}^{off}={{b}_{5,l}}+2\Re \left\{ \mathbf{u}_{l}^{H}{{\mathbf{t}}_{5,l}} \right\}-\mathbf{u}_{l}^{H}{{\mathbf{T}}_{5,l}}{{\mathbf{u}}_{l}},\tag{42}
\end{equation}
where several coefficients are defined as follows
\begin{align*}
& {{b}_{5,l}}=\log \left( 1+{{\alpha }_{2,l}} \right)-{{\alpha }_{2,l}},{{\mathbf{t}}_{5,l}}=\sqrt{1+{{a}_{2,l}}}\sqrt{{{p}_{l}}}{{\mathbf{g}}_{l}},\tag{43} \\ 
& {{\mathbf{T}}_{5,l}}={{\left| {{\beta }_{2,l}} \right|}^{2}}\left( \sum\limits_{l'=1}^{L}{{{p}_{l'}}{{\mathbf{g}}_{l'}}\mathbf{g}_{l'}^{H}}+{{\mathbf{H}}_{SI}}\mathbf{x}{{\left( {{\mathbf{H}}_{SI}}\mathbf{x} \right)}^{H}}+{{\mathbf{I}}_{{{N}_{r}}}}\sigma _{b}^{2} \right). 
\end{align*}

Based on the above transformation, the problem is reformulated as
\begin{equation}
{{\mathcal{P}}^{15}}:\underset{\mathbf{u}}{\mathop{\min }}\,\text{   }\sum\limits_{l=1}^{L}{\left( \mathbf{u}_{l}^{H}{{\mathbf{T}}_{5,l}}{{\mathbf{u}}_{l}}-2\Re \left\{ \mathbf{u}_{l}^{H}{{\mathbf{t}}_{5,l}} \right\}-{{b}_{5,l}} \right)}\tag{44}
\end{equation}

Problem ${{\mathcal{P}}^{15}}$ can be decomposed into $L$ independent sub-problems, each of which can be expressed as
\begin{equation}
\mathcal{P}_{l}^{16}:\underset{{{\mathbf{u}}_{l}}}{\mathop{\min }}\,\text{   }\mathbf{u}_{l}^{H}{{\mathbf{T}}_{5,l}}{{\mathbf{u}}_{l}}-2\Re \left\{ {{\mathbf{t}}_{5,l}}{{\mathbf{u}}_{l}} \right\}\tag{45}
\end{equation}

 Notice that Problem ${{\mathcal{P}}^{16}}$ is a unconstrained convex problem. By setting its derivative to zero, the optimal solution can be obtained as
follows
\begin{equation}
{{\mathbf{\hat{u}}}_{l}}={{\left( {{\mathbf{T}}_{5,l}} \right)}^{-1}}{{\mathbf{t}}_{5,l}},\forall l\in \mathcal{L}.\tag{46}
\end{equation}

\subsection {Optimize User Power and Local Computing Resources}
This section considers optimizing the transmission power and local computing resources of all CP-UEs when other variables are given, which is reformulated as
\begin{align*}
 {{\mathcal{P}}^{17}}:&\underset{\mathbf{f},\mathbf{p}}{\mathop{\max }}\,\text{    }\sum\limits_{k=1}^{K}{\tilde{R}_{k}^{com}}+\sum\limits_{l=1}^{L}{\left( \tilde{R}_{l}^{off}+\frac{R_{l}^{loc}}{B} \right)}\tag{47} \\ 
 \text{ }s.t.&\text{      }C2\text{ :  }{{r}^{tar}}\ge {{\Gamma }^{tar}},\tag{47a} \\ 
& \text{           }C\text{4 :  }T{{p}_{l}}+T\zeta f_{l}^{3}\le E_{l}^{\max },{{f}_{l}}\ge 0,{{p}_{l}}\ge 0.\tag{47b}
\end{align*}

Next, we rewrite the objective function and constraint C2 as
\begin{align*}
& \tilde{R}_{k}^{com}={{b}_{10,k}}-{{c}_{1,k}}\sum\limits_{l=1}^{L}{{{p}_{l}}{{b}_{11,k,l}}},\tag{48} \\ 
& \tilde{R}_{l}^{off}={{b}_{2,l}}+\sqrt{{{p}_{l}}}{{b}_{6,l}}-{{p}_{l}}{{b}_{7,l}},\tag{49} \\
&\hat{C}2:\sum\limits_{l=1}^{L}{{{p}_{l}}{{b}_{9,l}}}\le {{c}_{8}},\tag{50}
\end{align*}
where several coefficients are defined as follows
\begin{flalign*}
& {{b}_{10,k}}=\log \left( 1+{{\alpha }_{1,k}} \right)-{{\alpha }_{1,k}}+2\sqrt{1+{{\alpha }_{1,k}}}\Re \left\{ \beta _{1,k}^{*}{{\mathbf{h}}_{k}}{{\mathbf{w}}_{k}} \right\} \\ 
& -{{\left| {{\beta }_{1,k}} \right|}^{2}}\left( \sum\limits_{k'=0}^{K}{{{\left| {{\mathbf{h}}_{k}}{{\mathbf{w}}_{k'}} \right|}^{2}}}+\sigma _{k}^{2} \right),\tag{51} \\ 
& {{b}_{11,k,l}}={{\left| {{e}_{l,k}}+\mathbf{h}_{PU,k}^{H}\mathbf{\Phi }{{\mathbf{g}}_{PU,l}} \right|}^{2}}, \\ 
& {{b}_{2,l}}=\log \left( 1+{{\alpha }_{2,l}} \right)-{{\alpha }_{2,l}}\\
&-{{\left| {{\beta }_{2,l}} \right|}^{2}}\left( {{\left| \mathbf{u}_{l}^{H}{{\mathbf{H}}_{SI}}\mathbf{x} \right|}^{2}}+\left\| \mathbf{u}_{l}^{H} \right\|_{2}^{2}\sigma _{b}^{2} \right), \\ 
& {{b}_{6,l}}=2\Re \left\{ \sqrt{1+{{\alpha }_{2,l}}}\mathbf{u}_{l}^{H}{{\mathbf{g}}_{l}} \right\},{{b}_{7,l}}=\sum\limits_{l'=1}^{L}{{{\left| {{\beta }_{2,l'}} \right|}^{2}}{{\left| \mathbf{u}_{l'}^{H}{{\mathbf{g}}_{l}} \right|}^{2}}}, \\
&{{c}_{8}}=\left\| {{\mathbf{G}}_{s}}\mathbf{\Phi }{{\mathbf{G}}_{t}}\mathbf{x} \right\|_{2}^{2}-\sigma _{s}^{2}{{\Gamma }^{tar}},{{b}_{9,l}}={{\Gamma }^{tar}}\left\| {{\mathbf{g}}_{AU,l}} \right\|_{2}^{2}.
&
\end{flalign*}
Problem ${{\mathcal{P}}^{17}}$ can be formulated as
\begin{align*}
 {{\mathcal{P}}^{18}}:&\underset{\mathbf{f},\mathbf{p}}{\mathop{\max }}\,\text{    }\sum\limits_{l=1}^{L}{\left( {{b}_{6,l}}\sqrt{{{p}_{l}}}-{{p}_{l}}({{b}_{7,l}}+\sum\limits_{k=1}^{K}{{{c}_{1,k}}{{b}_{11,k,l}}})+\frac{{{f}_{l}}}{{{\varepsilon }_{l}}B} \right)}\tag{52} \\ 
\text{ }s.t.& \text{      }\bar{C}2\text{ :  }\sum\limits_{l=1}^{L}{{{p}_{l}}{{b}_{9,l}}}\le {{c}_{8}},\tag{52a} \\ 
& \text{            }C\text{4 :  }T{{p}_{l}}+T\zeta f_{l}^{3}\le E_{l}^{\max },{{f}_{l}}\ge 0,{{p}_{l}}\ge 0.\tag{52b} 
\end{align*}

The problem is concave in terms of $\mathbf{f}$ and $\mathbf{p}$, so it is a convex optimization problem and can be directly optimized with existing solvers.

\subsection {Algorithm Convergence, Optimality and Complexity Analysis}

The overall algorithm for solving problem ${{\mathcal{P}}^{0}}$ is summarized in Algorithm 1. After giving appropriate initial values, we iteratively update each subproblem until convergence. Note that the penalty parameter $\rho$ is reduced at each iteration to enforce satisfying the equality constraints. As the penalty parameter is gradually reduced, i.e., $\rho \to 0$, the solution to problem ${{\mathcal{P}}^{7}}$ eventually guarantees the satisfaction of the unit modulus constraint. And for any given $\rho$, the objective value achieved by problem ${{\mathcal{P}}^{7}}$ is an upper bound on the objective value achieved by problem ${{\mathcal{P}}^{5}}$. By alternately solving problem ${{\mathcal{P}}^{7}}$, the upper bound can be gradually tightened. Since each subproblem of problem ${{\mathcal{P}}^{7}}$ obtains an optimal solution, the objective function (27) is monotonically non-increasing, and the solution obtained by alternative optimization can be guaranteed to converge to the stable point of problem ${{\mathcal{P}}^{5}}$. Furthermore, we can find that problem ${{\mathcal{P}}^{2}}$ is solved by updating $\phi ,\mathbf{w},\mathbf{u},\left\{ \mathbf{f},\mathbf{p} \right\}$ alternately and each subproblem can converge to a stationary points. Note that the objective value (13) is non-decreasing over the iterations and that any limit point of any optimization variable is a stationary point of the original optimization problem ${{\mathcal{P}}^{2}}$. In addition, due to the transmit power budget and energy consumption budget, the upper bound of the objective value (13) is limited. Therefore, Algorithm 1 can converge to a stable point and a local optimal solution.

Next, we analyze the computational complexity of the proposed algorithm. As shown in Algorithm 1, the computational complexity mainly comes from the update of $\phi ,\mathbf{w},\mathbf{u},\left\{ \mathbf{f},\mathbf{p} \right\}$. We assume that the usual interior point method is used to solve these convex subproblems. Therefore, the complexity of updating $\phi$ is $\mathcal{O}\left\{ {{M}^{3.5}} \right\}$. The computational complexity of updating $\mathbf{w}$ is $\mathcal{O}\left( {{K}^{4.5}}\left( {{N}_{t}}^{4.5}+1 \right)\log \left( 1/\backepsilon   \right) \right)$, where $\backepsilon$ is the solution accuracy \cite{ref35}. And the computational complexity of updating $\left\{ \mathbf{f},\mathbf{p} \right\}$ is $\mathcal{O}\left\{ {{L}^{3.5}} \right\}$. Therefore, the total complexity of Algorithm 1 is $\mathcal{O}\left( {{I}_{o}}\left( {{M}^{3.5}}+{{K}^{4.5}}\left( {{N}_{t}}^{4.5}+1 \right)\log \left( 1/\backepsilon   \right)+{{L}^{3.5}} \right) \right)$, where ${{I}_{o}}$ represents the number of iterations required.

\begin{table}
	\begin{center}
		\caption{SYSTEM PARAMETERS.}
		\label{tab1}
		\begin{tabular}{ c | c | c}
			\hline
			Variable&Description & Value\\
			\hline
			${{N}_{t}}$ & The number of transmit antennas at BS &  4\\
			\hline
			${{N}_{r}}$ & The number of receive antennas at BS &  4\\ 
			\hline
			$M$ & The number of passive elements & 50\\
			\hline
			${{M}_{a}}$ & The number of active elements & 10\\
			\hline
			$K$ & The number of CM-UEs & 2\\
			\hline
			$L$ & The number of CP-UEs & 2\\
			\hline 
			$B$ & Bandwidth &  1 MHz\\
			\hline 
			${{P}_{b}}$ & Maximum transmit power of BS &  30 dBm\\
			\hline 
			${{\varepsilon }_{l}}$ & Complexity of computation tasks of CP-UE& 1000 cycles/bit\\
			\hline 
			$E_{m}^{\max }$ &The maximum energy budget for CP-UE $m$ &  0.01 J\\
			\hline 
			$\zeta$ & Constants related to hardware architecture & ${{10}^{-26}}$\\
			\hline 
			${{\Gamma }^{tar}}$ & Predefined target sensing threshold & 7 dB\\
			\hline 
			${{R}_{0,l}}$=${{R}_{0}}$ & The backhaul data rate & 100 $Mbps$\\
			\hline 
			$V$ & The number of file & 1000\\
			\hline 
			$F$ & The storage capacity  & ${{10}^{6}}$\\
			\hline 
			${{q}_{v}}$ & The length of files  & ${{10}^{5}}$\\
			\hline 
			$\epsilon $ & Skewness factor  & 1.4\\
			\hline  
		\end{tabular}
	\end{center}
\end{table}

\section{Simulation Results}
In this section, we give the simulation results to verify the performance of the considered systems. First, we model BS TX-IRS RE links, BS RX-IRS RE links, IRS RE-CP UEs links, IRS SE-CP UEs links, and IRS RE-CM UEs links as Rician distribution with a Ricean Rician of 3dB. The direct link between CP-UEs and CM-UEs is modeled as Rayleigh fading \cite{ref36}. For all channels, we adopt the typical distance-dependent path loss model \cite{ref20}. The large-scale fading model is denoted as $PL\left( d \right)=\Lambda {{\left( \frac{d}{{{d}_{0}}} \right)}^{-\eta }}$, where $\Lambda =-30\text{dB}$ represents the path loss at the reference distance ${{d}_{0}}=1m$, and $d$ and $\eta $ represent the propagation distance and path loss exponent, respectively. The path loss exponents between BS-IRS, IRS-UE, IRS-Target, and CP UE-CM UE are ${{\eta }_{BR}}=2.2$, ${{\eta }_{RU}}=2.5$, ${{\eta }_{RT}}=2.2$ and ${{\eta }_{MP}}=3.9$ respectively. For the SI channel, according to \cite{ref37}, each entry of model ${{\mathbf{H}}_{SI}}\in {{\mathbb{C}}^{{{N}_{r}}\times {{N}_{t}}}}$ is ${{\left[ {{\mathbf{H}}_{SI}} \right]}_{t,r}}=\sqrt{\varpi _{t,r}^{SI}}{{e}^{-j2\pi \frac{{{d}_{t,r}}}{\lambda }}}$, where ${{d}_{t,r}}>0$ represents the distance between the $t$-th transmit antenna and the $r$-th receive antenna and $\varpi _{t,r}^{SI}>0$ represents the residual SI power. For simplicity, we set ${{\varpi }^{SI}}=\varpi _{t,r}^{SI}=-110\text{dB}$ and let ${{e}^{-j2\pi \frac{{{d}_{t,r}}}{\lambda }}}$ be a unit modulus variable with random phase for all transceiver antenna pairs $\left( t,r \right)$ \cite{ref38}. 

Several specific system parameters are shown in Table 1. In addition, we assume that BS and IRS are deployed at (-50m, 0) and (0, 6m). All users are randomly distributed in $\left( {{d}_{x}}\text{m},{{d}_{y}}\text{m} \right)$, where ${{d}_{x}}\in [10,40]$ and ${{d}_{y}}\in [0,1]$. Assume that the radar target is 3m away from the IRS and $\theta ={{40}^{\circ }}$. The noise power at the BS, each CM-UE, and IRS is set to $\sigma _{b}^{2}=\sigma _{k}^{2}=\sigma _{s}^{2}=-90\text{dBm},\forall k$.
 
In order to verify the performance of the proposed system, we also include the following solutions for comparison:
\begin{itemize}
	\item{Full offloading: Each CP-UE performs computation offloading only through the IRS.}
	\item{Fixed phase shift: We assume that the IRS adopts a fixed phase shift determined by the maximum channel gain.}
	\item{HD: Assume that BS uses HD mode, where CM-UEs reception and CP-UEs transmission are implemented in two orthogonal time slots of equal duration. Therefore, the CCI of uplink offloading signals to CM-UEs and SI at the BS do not exist.}
	\item{Optimizing caching: In each implementation, the caching strategy is optimized by solving problem ${{\mathcal{P}}^{1}}$.}
	\item{Random caching strategy: The probability of BS local random caching depends on the popularity of the cached data, i.e., if the data is more popular, it is more likely to be cached.}
	\item{Without caching: Both computation offloading and local computation can be performed without caching.}
\end{itemize}

\begin{figure}[!t]
	\centering
	\includegraphics[width=3.5in]{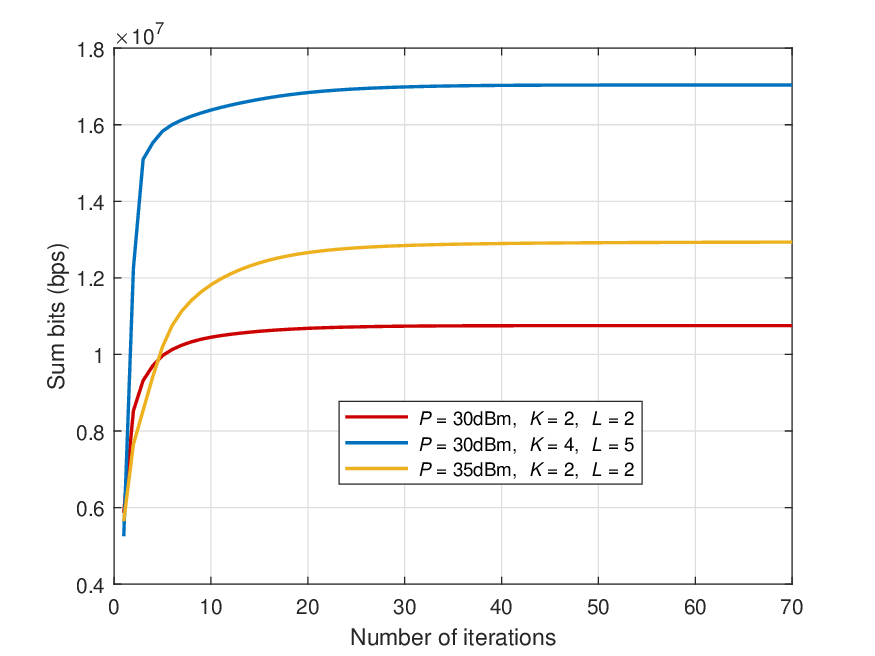}
	\captionsetup{justification=centering} 
	\caption{Sum bits versus the number of iterations.}
	\label{fig2}
\end{figure} 
\begin{figure}[!t]  
	\centering
	\includegraphics[width=3.5in]{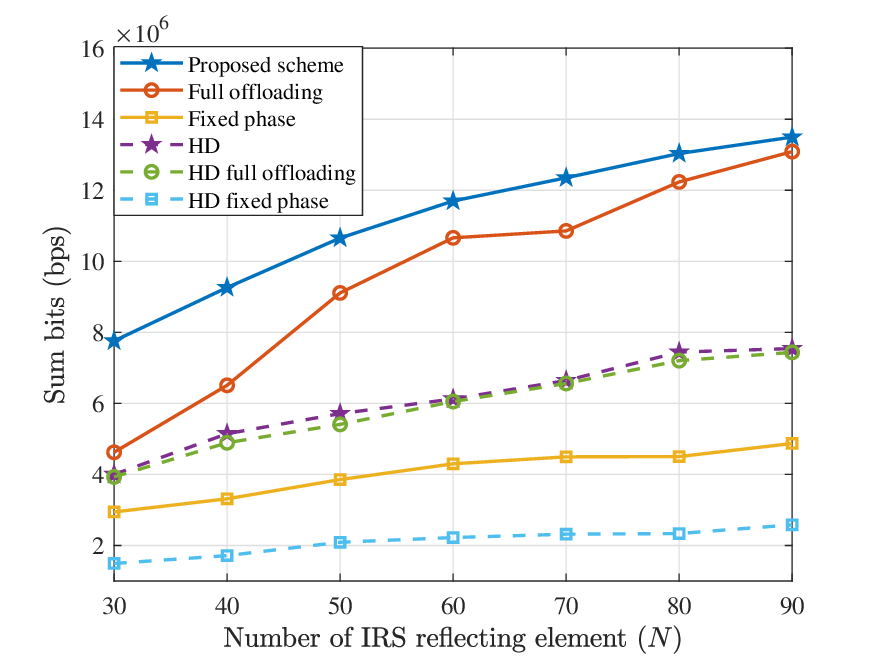}
	\captionsetup{justification=centering}
	\caption{Sum bits versus the number of IRS reflection elements $N$.}
	\label{fig3} 
\end{figure}

\begin{figure}[t]   
	\centering
	\includegraphics[width=3.5in]{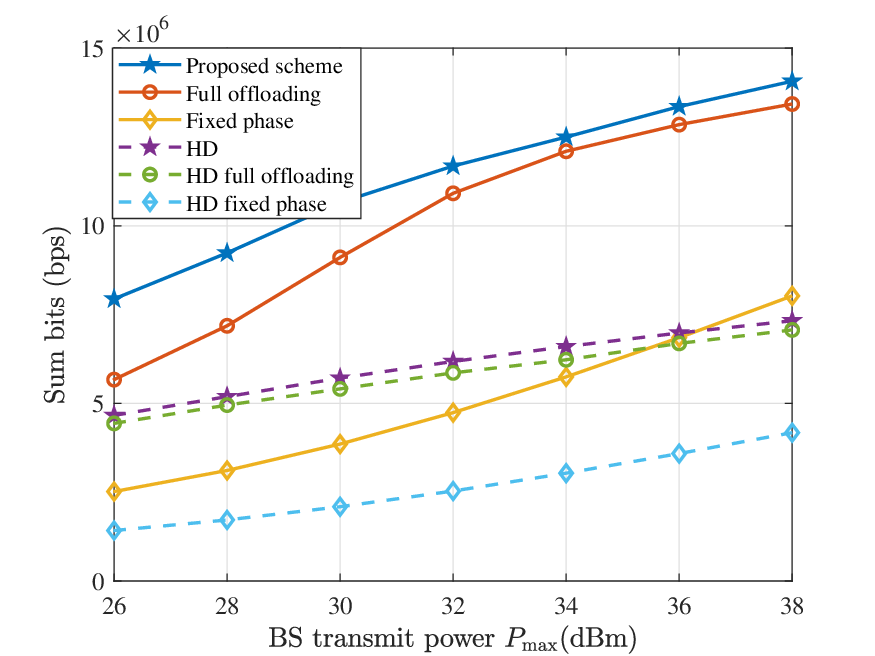}
	\captionsetup{justification=centering}
	\caption{Sum bits versus the BS transmit power ${{P}_{BS}}$.}
	\label{fig4} 
\end{figure} 
\begin{figure}[!t]  
	\centering
	\includegraphics[width=3.5in]{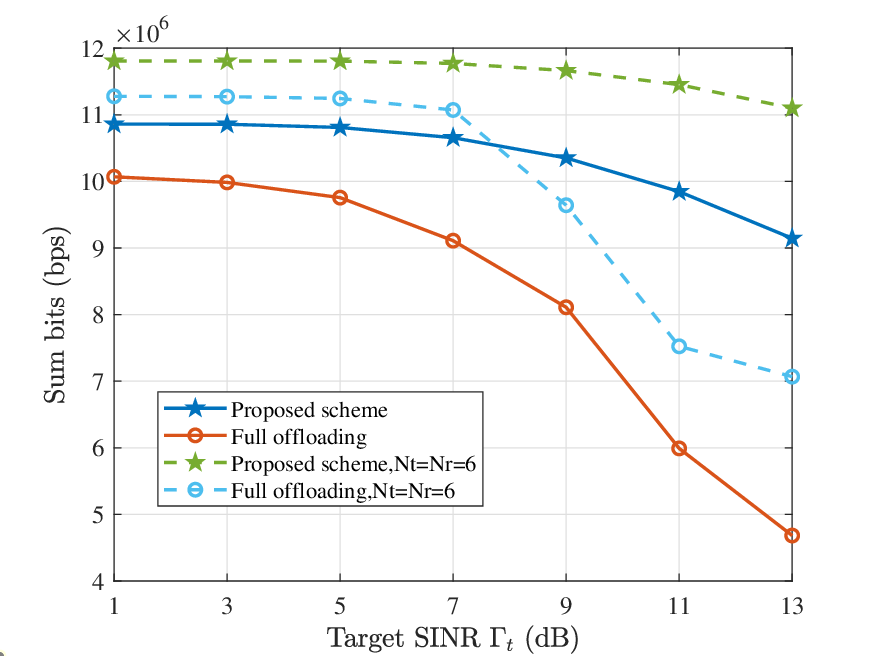}
	\captionsetup{justification=centering} 
	\caption{Sum bits versus the target sensing requirement ${{\Gamma }^{tar}}$.}
	\label{fig5} 
\end{figure}

Fig. 2 shows the convergence of the proposed algorithm. We present different settings for the relationship between the achievable sum bits and the number of iterations to demonstrate the convergence performances of Algorithm 1. We observe that after 30 iterations, the achievable sum bits trend to stablize, verifying the fast convergence and effectiveness of the proposed algorithm. In addition, it can be seen that as the number of users increases, a higher sum bits can usually be obtained.

Fig. 3 depicts the relationship between the sum bits of the proposed scheme and the benchmark scheme and the number of IRS reflection elements $N$. It can be seen that by increasing the number of IRS reflection elements, the sum bits increase under all schemes. This is because that more reflective elements can tailor favorable channels with more degrees of freedom, providing greater passive beamforming gain for all users. Furthermore, our proposed algorithm outperforms both full offloading and fixed phase shift in both FD and HD systems. This highlights the optimality of partial offloading and that by optimizing the phase shift of the IRS, a more favorable propagation environment can be created. In addition, the performance of the FD scheme, i.e., the proposed scheme, is significantly better than that of the traditional HD scheme, this is because in the same time, the HD scheme can fully utilize the degrees of freedom introduced by multiplexing uplink and downlink users on the same spectrum resources, thereby improving system performance.

Fig. 4 depicts the relationship between the sum bits and the maximum transmission power ${{P}_{BS}}$ of the BS for the proposed scheme and the benchmark scheme. As expected, the sum bits of all schemes increases monotonically with increasing ${{P}_{BS}}$. A larger ${{P}_{BS}}$ provides higher beamforming gain for signal transmission, thereby improving system performance. Furthermore, we observe that the proposed scheme outperforms all other schemes. In fact, compared with other schemes, through the joint optimization of $\phi ,\mathbf{w},\mathbf{u},\mathbf{f}$ and $\mathbf{p}$, this resource allocation scheme can be significantly improved. In contrast, for the full offloading scheme, as increases ${{P}_{BS}}$, both CCI and the remaining SI become more severe, which affects the performance of the system. For the FD scheme, although orthogonal transmission avoids CCI and residual SI, the resulting strictly suboptimal use of uplink and downlink time resources leads to a degradation of system performance.

\begin{figure}[!t]  
	\centering
	\includegraphics[width=3.5in]{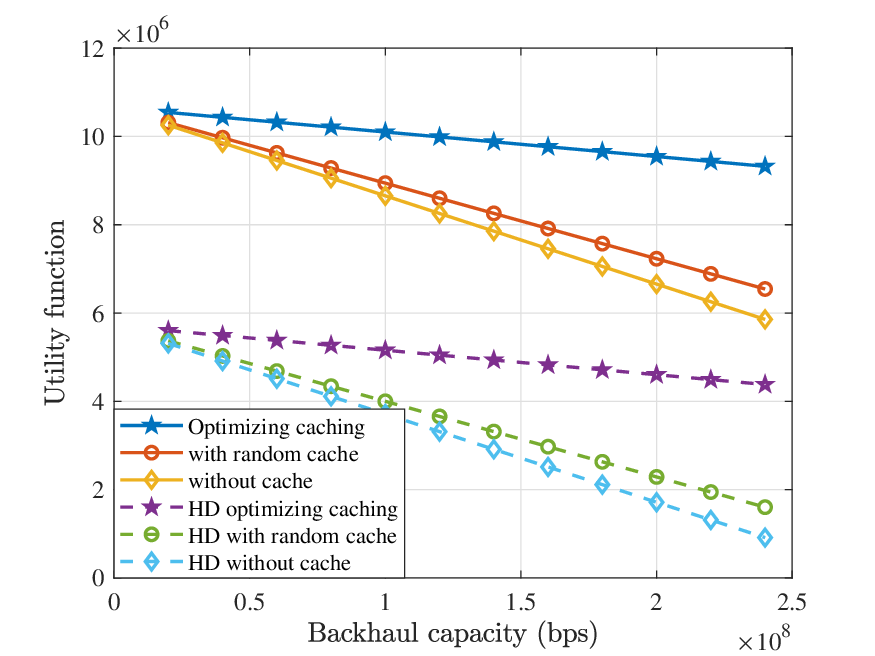}
	\captionsetup{justification=centering} 
	\caption{System utility function versus the backhaul data rate ${{R}_{0}}$.}
	\label{fig6} 
\end{figure} 
\begin{figure}[!t]  
	\centering
	\includegraphics[width=3.5in]{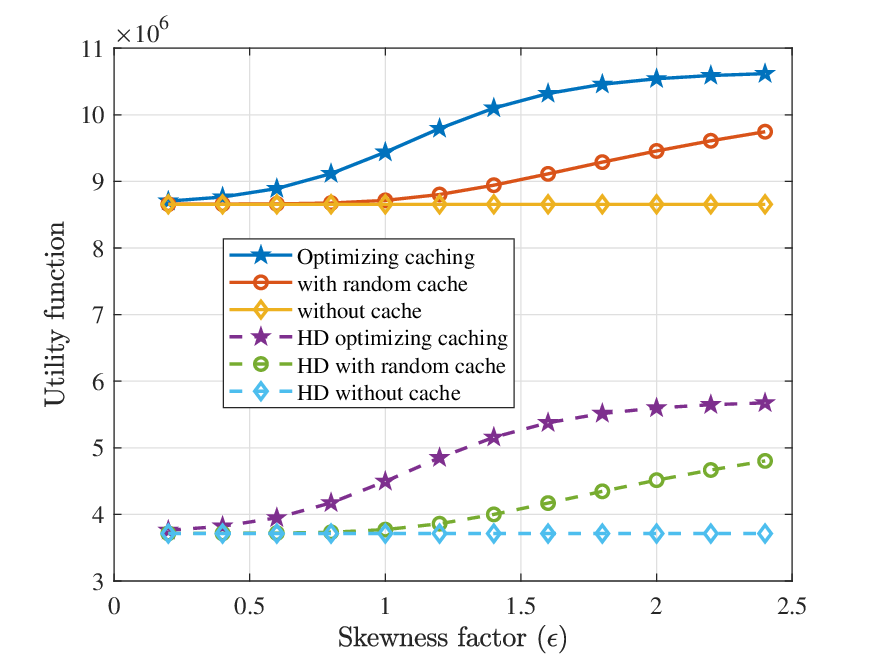}
	\captionsetup{justification=centering} 
	\caption{System utility function versus the skewness factor $\epsilon $.}
	\label{fig7} 
\end{figure} 

Fig. 5 shows the relationship between the sum bits of the proposed scheme and the baseline scheme based on different target sensing requirement ${{\Gamma }^{tar}}$. We can observe that the sum bits of all solutions decreases as ${{\Gamma }^{tar}}$ increases. This is because when ${{\Gamma }^{tar}}$ increases, the system needs to reduce the communication rate to utilize more transmission power to ensure the sensing demand, which leads to a decrease in system performance. Furthermore, we can see that the system performance improves as the number of BS antennas increases. This is due to the extra degrees of freedom provided by the additional antennas, which provide higher beamforming resolution for both CM-UE reception and CP-UE transmission, leading to higher system performance.

Fig. 6 shows the relationship between the system utility function and the backhaul data rate ${{R}_{0}}$. It can be observed that as ${{R}_{0}}$ increases, the system utility functions of all schemes decrease, which can be explained by the fact that the increase in the backhaul data rate leads to an increase in the total cost of content delivery, thus resulting in a decrease in the system utility function. In addition, the optimal caching scheme outperforms the random caching and without caching schemes in both FD and HD systems.

Fig. 7 shows the relationship between the system utility function and the skewness factor $\epsilon $ in the Zipf distribution function. As can be observed from the figure, as $\epsilon $ increases, the system utility functions of all schemes first gradually show a linear increase trend. When $\epsilon $ is larger, the growth of the optimal caching gradually slows down. In addition, the optimal caching scheme outperforms the random caching and without caching schemes in both FD and HD systems. And the without scheme is not affected by $\epsilon $ and remains unchanged.

\section{Conclusion}
In this paper, we studied the resource allocation problem in an IRS-assisted FD ISCC system. We proposed an optimization problem to maximize the system utility function while satisfying the worst-case radar SINR constraint, transmission power budget, transmission energy budget, storage capacity budget, and IRS reflection coefficient unit modulus constraint. An efficient algorithm was developed to solve this non-convex problem. Finally, we discussed the complexity of the algorithm and verified the effectiveness of the algorithm and its advantages over other benchmark algorithms through simulation results. Inspired by this work, we will further explore the application scenarios of IRS in ISCC systems, including IRS-assisted FD ISCC systems with hardware impairments and discrete phase shifts.

\end{document}